\documentclass[12pt]{amsart}
\usepackage{graphicx}

\input{epsf}

\theoremstyle{definition}

\theoremstyle{remark}

\numberwithin{equation}{section}
\begin{document}

\title[SAE in SUSYQM]
{Self-adjoint extensions and SUSY breaking in Supersymmetric Quantum Mechanics}%
\author{H. Falomir and P.\ A.\ G.\ Pisani}%
\address{IFLP (CONICET) - Departamento de F{\'\i}sica,
Fac.\ de Ciencias Exactas, UNLP,
C.C.\ 67, (1900) La Plata, Argentina}%
\email{falomir@fisica.unlp.edu.ar,
pisani@fisica.unlp.edu.ar}%

%\thanks{}%
%\subjclass{}%
%\keywords{}%

\date{\today}%
%\dedicatory{}%
%\commby{}%
% ----------------------------------------------------------------
\begin{abstract}
We consider the self-adjoint extensions (SAE) of the symmetric
supercharges and Hamiltonian for a model of SUSY Quantum Mechanics
in $\mathbb{R}^+$ with a singular superpotential. We show that
only for two particular SAE, whose domains are scale invariant,
the algebra of $N=2$ SUSY is realized, one with manifest SUSY and
the other with spontaneously broken SUSY. Otherwise, only the
$N=1$ SUSY algebra is obtained, with spontaneously broken SUSY and
non degenerate energy spectrum.



\end{abstract}
\maketitle
% ----------------------------------------------------------------
\section{Introduction}

Supersymmetry (SUSY)
\cite{Gelfand,Ramond,Neveu,Volkov,W-Z,Fayet,Witten,Salomonson,Cooper,BCD}
gives desirable features to quantum field theories, like an
improved ultraviolet behavior, but also predicts superpartner
states with degenerate mass which are not observed experimentally.
Therefore, this symmetry is expected to be spontaneously
(dynamically) broken.

Several schemes have been developed to try to solve the SUSY
breaking problem, including the idea of non-perturbative breaking
by instantons. In this context, the simplest case of SUSY Quantum
Mechanics (SUSYQM) was introduced by Witten \cite{Witten} and
Cooper and Freedman \cite{Cooper}.

When considering these models, several authors have suggested that
singular potentials could break SUSY through nonstandard
mechanisms, being responsible for non degeneracy of energy levels
and negative energy eigenstates \cite{JR,Casa,Imbo,Roy,Pani,CKS}.

In particular, Jevicki and Rodrigues \cite{JR} have considered the
singular superpotential $W(x)=g/x -x$. Based on the square
integrable solutions of a differential operator related to the
Hamiltonian of this system \cite{Lat} they concluded that, for a
certain range of the parameter $g$, SUSY is broken with a negative
energy ground state.

However, they have not considered if all these functions
correspond to eigenvectors of a unique self-adjoint Hamiltonian.
As is well known, the quantum dynamics is given by a unitary
group, and it follows from Stone's theorem \cite{R-S} that the
Hamiltonian, which is the infinitesimal generator of this group,
must be self-adjoint.

\smallskip

Later, Das and Pernice \cite{DP} have reconsidered this problem in
the framework of a SUSY preserving regularization of the singular
superpotential, finding that SUSY is recovered manifestly at the
end, when the regularization is removed. They conclude that SUSY
is robust at short distances (high energies), and the
singularities that occur in quantum mechanical models are unlike
to break SUSY.

\bigskip

In the present article we would like to address this controversial
subject by studying the self-adjoint extensions of the Hamiltonian
defined by the singular superpotential $W(x)=g/x -x$ with $x\in
\mathbb{R}^+$. This will be done by studying the self-adjoint
extensions of the symmetric supercharges, and by considering the
possibility of realizing the algebra of SUSY in a dense subspace
of the Hilbert space.

We will show that there is a range of values of $g$ for which the
supercharges admit a one-parameter family of self-adjoint
extensions, corresponding to a one-parameter family of
self-adjoint extensions of the Hamiltonian. We will show that only
for two particular self-adjoint extensions, whose domains are
scale invariant, the algebra of $N=2$ SUSY can be realized, one
with manifest SUSY and the other with spontaneously broken SUSY.
For other values of this continuous parameter, only the $N=1$ SUSY
algebra is obtained, with spontaneously broken SUSY and non
degenerate energy spectrum.

\smallskip

We should mention that self-adjoint extensions of  supercharges
and Hamiltonian for the SUSYQM of the free particle with a point
singularity in the line and the circle have been considered in
\cite{UT,ChFT,TFCh,NST}, where $N=1,2$ realization of SUSY are
described. They have also been considered in the framework of the
Landau Hamiltonian for two-dimensional particles in nontrivial
topologies in \cite{Allen} (see also \cite{FP1}).

\bigskip

Let us remark that, given a superpotential $W(x)$, one gets a {\em
formal} expression for the Hamiltonian (and also for the
supercharges) as a symmetric differential operator $H$ defined on
a subspace of sufficiently smooth square-integrable functions. The
theory of {\em deficiency indices} of von Neumann \cite{R-S} gives
the basic criterion for the existence of {\em self-adjoint
extensions} of this operator. In the case where there is only one
self-adjoint extension, $H$ is {\em essentially self-adjoint} and
its {\em closure} \cite{R-S} represents the true Hamiltonian of
the system. But if there are several self-adjoint extensions of
$H$, they usually differ by the {\em physics} they describe. In
this case, the election of a Hamiltonian among the self-adjoint
extensions of $H$ is not just a mathematical technicality. Rather,
additional physical information is required  to select the correct
one, that which describes the true properties of the system.

\bigskip

The structure of the paper is as follows: In the next Section we
present the problem to solve. In Section \ref{Sec-adjointQ+} we
study the adjoint operator of the supercharge, whose properties
are needed to determine the supercharge self-adjoint extensions.
This is done in Section \ref{Sec-SAEQ+}, where the self-adjoint
extensions of the Hamiltonian are also determined. In Section
\ref{Discussion} we consider the possibility of realizing the
algebra of the supersymmetry on the Hamiltonian domain of
definition, and state our conclusions. In Appendix \ref{closure}
we treat some technicalities related to the closure of the
symmetric supercharge and in Appendix \ref{spectral-functions} we
consider the graded partition function and the Witten index of the
Hamiltonian, and the spectral asymmetry of the supercharge.

\section{Setting of the problem}\label{setting}

The Hamiltonian of a supersymmetric one-dimensional system can be
written as
\begin{equation}\label{Ham-QQ}
  H=\{ \mathcal{Q}, \widetilde{\mathcal{Q}} \}_+,
\end{equation}
where the supercharges
\begin{eqnarray}\label{Q-Qtilde}
    \mathcal{Q}=\left(\begin{array}{cc}
            0&0\\
            A&0\
        \end{array}\right),\quad
    \widetilde{\mathcal{Q}}=\left(\begin{array}{cc}
            0&\widetilde{A}\\
            0&0\
        \end{array}\right)
\end{eqnarray}
are nilpotent operators,
\begin{equation}\label{nilpot}
    \mathcal{Q}^2=\widetilde{\mathcal{Q}}^2=0,
\end{equation}
which commute with the Hamiltonian.

In eq.\ (\ref{Q-Qtilde}),
\begin{equation}\label{A-Atilde}
  A=\displaystyle{\frac{1}{\sqrt{2}}\left(-\frac{d}{dx}+W(x)\right)}
  \quad {\rm and} \quad
  \widetilde{A}=
  \displaystyle{\frac{1}{\sqrt{2}}\left(\frac{d}{dx}+W(x)\right)}
\end{equation}
are differential operators defined on a suitable dense subspace of
functions where the necessary compositions of operators in Eqs.\
(\ref{Ham-QQ}) and (\ref{nilpot}) are well defined, and $W(x)$ is
the superpotential.

\smallskip

In this Section we will consider a quantum mechanical system
living in the half line $\mathbf{R^+}$, subject to a
superpotential given by
\begin{equation}\label{W}
  W(x)=\frac{g}{x}-x
\end{equation}
for $x>0$ and $g$ real. The two differential operators defined in
(\ref{A-Atilde}) take the form
\begin{eqnarray}
    A=\frac{1}{\sqrt{2}}\left(-\frac{d}{dx}+\frac{g}{x}-x\right),
    \label{superA}\\
    \widetilde{A}=\frac{1}{\sqrt{2}}\left(\frac{d}{dx}+
    \frac{g}{x}-x\right).\label{superAt}
\end{eqnarray}

\smallskip

Let us now introduce an operator $Q_{+}$, defined on the dense
subspace of (two component) functions with continuous derivatives
of all order and compact support not containing the origin,
$\mathcal{D}(Q_{+})=\mathcal{C}^{\infty}_{0}(\mathbf{R^+\backslash
\{0\}})$, over which its action is given by
\begin{equation}\label{P-def}
    Q_{+} \Psi=\left(\begin{array}{cc}
            0&\widetilde{A}\\
            A&0\
        \end{array}\right)\left(\begin{array}{c}
          \psi_1 \\
          \psi_2 \\
        \end{array}\right).
\end{equation}
Notice that, within this domain, $Q_{+}$ can be identified with
\begin{equation}
    Q_{+}=\widetilde{\mathcal{Q}}+\mathcal{Q},
\end{equation}
while its square (which is well defined) satisfies
\begin{equation}\label{susy-ham}
    Q_{+}^2=\{\mathcal{Q},\widetilde{\mathcal{Q}}\}_+=H=\left(\begin{array}{cc}
            H_+&0\\
            0&H_-\
        \end{array}\right),
\end{equation}
where $H$ is the Hamiltonian of the system, with
$H_+=\widetilde{A}A$ and $H_-=A\widetilde{A}$.

It can be easily verified that $Q_{+}$ so defined is a symmetric
operator, but it is neither self-adjoint nor even closed.
Consequently, we must look for the self-adjoint extensions of
$Q_{+}$.

\smallskip

Within the same domain, a linearly independent combination of
supercharges leads to the operator
\begin{equation}\label{Q-}
    Q_{-}=i(\widetilde{\mathcal{Q}}-\mathcal{Q}),
\end{equation}
which is also symmetric and satisfies that $Q_{-}^2 = H$, and
$\left\{ Q_{+},Q_{-} \right\}_+=0$. Since it can be obtained from
$Q_{+}$ through a unitary transformation given by
\begin{equation}\label{PP}
  Q_{-}= e^{i \sigma_3 \pi/4} Q_{+}e^{-i \sigma_3 \pi/4},\ {\rm with}\
  \sigma_3 = \left(\begin{array}{cc}
    1 & 0 \\
    0 & -1 \
  \end{array}\right),
\end{equation}
the following analysis will be carried out only for $Q_{+}$, and
it will extend immediately to $Q_{-}$.

\smallskip

Notice that, given a self-adjoint extension of $Q_{+}$ (which, in
particular, is a closed and densely defined operator \cite{R-S}),
its {\em square} gives the corresponding self-adjoint extension of
the Hamiltonian $H$ in Eq.\ (\ref{susy-ham}), by virtue of a
theorem due to von Neumann \cite{teor-vN}.

The first step in the construction of the self-adjoint extensions
of $Q_{+}$ consists in the determination of its adjoint,
$Q_{+}^\dagger$, which will be done in the next Section.

\section{The adjoint operator $Q_{+}^\dagger$ \label{Sec-adjointQ+}}

In this Section we will determine the domain of definition of
$Q_{+}^\dagger$, and its spectrum. In particular, we are
interested in the {\it deficiency subspaces} \cite{R-S} of $Q_{+}$
(the null subspaces of $(Q_{+}^\dagger \mp i)$),
\begin{equation}\label{defi-sub}
  \mathcal{K}_\pm
    := {\rm Ker}(Q_{+}^\dagger \mp i ),
\end{equation}
which determine the self-adjoint extensions of $Q_{+}$.

\subsection{Domain of $Q_{+}^\dagger$}

A (two component) function $\Phi$ belongs to the domain of
$Q_{+}^\dagger$,
\begin{equation}\label{dom-adjoint}
  \Phi=\left( \begin{array}{c}
  \phi_1   \\
  \phi_2
\end{array} \right) \in \mathcal{D}(Q_{+}^\dagger) \subset
\mathbf{L_2}(\mathbb{R^+}),
\end{equation}
if $\left(\Phi, Q_{+}\Psi\right)$ is a linear continuous
functional of $\Psi$, for $\Psi \in \mathcal{D}(Q_{+})$. This
requires the existence of a function
\begin{equation}
  \Omega=\left(
    \begin{array}{c}
  \chi_1   \\
  \chi_2
    \end{array}\right)  \in
    \mathbf{L_2}(\mathbb{R^+})
\end{equation}
such that
\begin{equation}\label{adjoint}
  \left(\Phi,Q_{+}\Psi\right)=\left(\Omega,\Psi\right), \ \forall \Psi
  \in \mathcal{D}(Q_{+}).
\end{equation}
Such $\Omega$ is uniquely determined, since $\mathcal{D}(Q_{+})$
is a dense subspace. Then, for each $\Phi \in
\mathcal{D}(Q_{+}^\dagger)$, the action of $Q_{+}^\dagger$ is
defined by $Q_{+}^\dagger \Phi := \Omega$. Notice that
$\mathcal{D}(Q_{+}) \subset \mathcal{D}(Q_{+}^\dagger)$, since
$Q_{+}$ is symmetric.

\bigskip

We will now determine the properties of the functions in
$\mathcal{D}(Q_{+}^\dagger)$, and the way $Q_{+}^\dagger $ acts on
them. In a distributional sense, Eq.\ (\ref{adjoint}) implies that
\begin{eqnarray}
    \displaystyle{
  -\phi_1'+\left({g\over x}-x\right)\phi_1=\sqrt{2} \, \chi_2},
     \label{deriv-prim} \\ \displaystyle{
    \phi_2'+\left({g\over x}-x\right)\phi_2=\sqrt{2}\, \chi_1}
    \label{deriv-prim-2},
\end{eqnarray}
which shows that $\Phi'(x)$ is a regular (locally integrable)
distribution. This implies that $\Phi(x)$ is an absolutely
continuous function for $x>0$.

Therefore, the domain of $Q_{+}^\dagger$ consists on those
(square-integrable) absolutely continuous functions such that the
left hand sides in Eqs.\ (\ref{deriv-prim}) and
(\ref{deriv-prim-2}) are also square-integrable functions on the
half-line:
\begin{equation}\label{dom-P+}\begin{array}{c}
  \mathcal{D}(Q_{+}^\dagger)=\{
  \Phi \in AC(\mathbb{R^+}\backslash\{0\}) \cap
  \mathbf{L_2}(\mathbb{R^+}):
  \\ \\
   A\,\phi_1,\, \widetilde{A}\,\phi_2 \in  \mathbf{L_2}(\mathbb{R^+})
  \}.
\end{array}
\end{equation}
Consequently, an integration by parts in the left hand side of
Eq.\ (\ref{adjoint}) is justified, and we conclude that the action
of $Q_{+}^\dagger$ on $\Phi \in  \mathcal{D}(Q_{+}^\dagger)$ also
reduces to the application of the differential operator
\begin{equation}\label{P+-def}
    Q_{+}^\dagger \Phi=\left(\begin{array}{cc}
            0&\widetilde{A}\\
            A&0\
        \end{array}\right)\left(\begin{array}{c}
          \phi_1 \\
          \phi_2 \\
        \end{array}\right).
\end{equation}

\subsection{Spectrum of $Q_{+}^\dagger$} \label{spectrum}

We now consider the eigenvalue problem for $Q_{+}^\dagger$,
\begin{equation}\label{P-auto}
  Q_{+}^\dagger \Phi_\lambda = \lambda \Phi_\lambda,
\end{equation}
or equivalently
\begin{equation}\label{eigen-problem}
  A\,\phi_1=\lambda\, \phi_2, \quad \widetilde{A}\,\phi_2=\lambda\,
  \phi_1,
\end{equation}
with
\begin{equation}\label{phi-eigen}
  \Phi_\lambda = \left(
    \begin{array}{c}
      \phi_1 \\
      \phi_2
    \end{array}
\right) \in \mathcal{D}(Q_{+}^\dagger)
\end{equation}
 and $\lambda \in \mathbb{C}$.

From Eqs.\ (\ref{superA}), (\ref{superAt}) and
(\ref{eigen-problem}), it follows immediately that
$\Phi_\lambda'(x)$ is an absolutely continuous function. In fact,
the successive applications of $Q_{+}^\dagger $ on both sides of
Eq.\ (\ref{P-auto}) show that $\Phi_\lambda(x) \in
\mathcal{C}^{\infty}(\mathbb{R^+}\backslash \{0\})$, and Eq.\
(\ref{eigen-problem}) is just a system of ordinary differential
equations.

\bigskip

Replacing $\phi_2$ in terms of $\phi_1$ we get
\begin{eqnarray}
    -\frac{1}{2}\,\phi_1''+\frac{1}{2}
            \left\{\frac{g(g-1)}{x^2}+x^2-1-2g\right\}\phi_1=
            \lambda^2\, \phi_1,\label{sys1}\\
   \lambda\, \phi_2=\frac{1}{ \sqrt{2}}
    \left\{-\phi_1'+
    \left(\frac{g}{x}-x\right)\phi_1\right\}. \label{sys2}
\end{eqnarray}
Making the substitution
\begin{equation}\label{phi1dex}
    \phi_1(x)=x^g\, e^{{- {x^2/ 2}} }\, F(x^2)
\end{equation}
in Eq.\ (\ref{sys1}) we get the Kummer's equation \cite{A-S} for
$F(z)$,
\begin{equation}\label{kummer-eq}
    z\, F''(z)+(b-z)\,F'(z) -a\, F(z)=0,
\end{equation}
with
\begin{equation}\label{a-b}
  a=- \frac{\lambda^2}{2},\quad b=g+\frac{1}{2}.
\end{equation}

For any values of the parameters $a$ and $b$, equation
(\ref{kummer-eq}) has two linearly independent solutions
\cite{A-S} given by the Kummer's function
\begin{equation}\label{LI-sol-1}
  \begin{array}{c}
    y_1(z)=U(a,b,z)= \\ \\
    \displaystyle{ \frac{\pi}{\sin \pi b}\left\{
    \frac{M(a,b,z)}{\Gamma(1+a-b)\Gamma(b)} -
    z^{1-b}\, \frac{M(1+a-b,2-b,z)}{\Gamma(a)\Gamma(2-b)}
    \right\} },
  \end{array}
\end{equation}
and
\begin{equation}\label{LI-sol-2}
  y_2(z)= e^z \, U(b-a,b,-z).
\end{equation}
In Eq.\ (\ref{LI-sol-1}), $M(a,b,z)={_1F_1(a;b;z)}$ is the
confluent hypergeometric function.

Since for large values of the argument \cite{A-S}
\begin{equation}\label{U-asymp}
  U(a,b,z)=z^{-a}\left\{1+ \mathcal{O}(|z|^{-1})\right\},
\end{equation}
only $y_1(x^2)$ leads to a function $\phi_1(x) \in
\mathbf{L_2}(1,\infty)$ when replaced in Eq.\ (\ref{phi1dex}),
while $y_2(x^2)$ should be discarded.

Therefore, we get
\begin{equation}\label{phi1-U}
  \phi_1(x)= x^g\, e^{{- {x^2/ 2}} }\,
  U(-\frac{\lambda^2}{2}, g+\frac{1}{2}, x^2).
\end{equation}
On the other hand, replacing Eq.\ (\ref{phi1-U}) in Eq.\
(\ref{sys2}), it is straightforward to show that \cite{A-S}
\begin{equation}\label{phi2-U}
    \phi_2(x)=-\frac{\lambda}{\sqrt{2}}\, x^{g+1}\, e^{-{x^2}/{2}}
        \, U(1-\frac{\lambda^2}{2};g+\frac{3}{2},x^2),
\end{equation}
which is also in $\mathbf{L_2}(1,\infty)$.

In order to determine the spectrum of $Q_{+}^\dagger$, we must now
consider the behavior of
    $\Phi_\lambda(x) = \left(\begin{array}{c}
    \phi_1(x) \\
    \phi_2(x)
    \end{array}\right)$
near the origin. From Eq.\ (\ref{LI-sol-1}), and the small
argument expansion of Kummer's functions (see \cite{A-S}, page
508), one can straightforwardly show that three cases should be
distinguished, according to the values of the coupling $g$:

\smallskip

\begin{itemize}

\item{If {$g\geq 1/2$}, it can be seen that $\Phi_\lambda(x)\notin
\mathbf{L_2}(0,1)$ unless $-\lambda^2/2 = -n$, with
$n=0,1,2,\dots$ In this case, taking into account that $U(-n, b,
z)$ reduces to a Laguerre polynomial (of degree $n$ in $z$),
\begin{equation}\label{U-poly}
    U(-n, b, z) = (-1)^n \, n! \,
    L_n^{(b-1)}(z),
\end{equation}
we have $\phi_1(x)\sim x^g$ and $\phi_2(x)\sim x^{g+1}$ for
$0<x\ll 1$ (Notice that the square-integrability of $\phi_{1}(x)$
and $\phi_{2}(x)$ on $\mathbb{R}^{+}$ is guaranteed by the
decreasing exponentials in Eqs.\ (\ref{phi1-U}) and
(\ref{phi2-U})). Therefore, in this region $Q_{+}^\dagger$ has a
symmetric real spectrum given by the (degeneracy one) eigenvalues
\begin{equation}\label{spectrum-1}
  \lambda_{0}=0,\quad \lambda_{\pm,n}= \pm \sqrt{2 n},\quad n=1,2,3,\dots
\end{equation}
corresponding to the eigenfunctions
\begin{equation}\label{eigenfunctions-1-0}
  \Phi_{0}=x^g\, e^{{- {x^2/ 2}} }
  \left(\begin{array}{c}
    1 \\0
  \end{array}\right),
\end{equation}
and
\begin{equation}\label{eigenfunctions-1}
  \Phi_{\pm,n}=(-1)^n \, n! \,x^g\, e^{{- {x^2/ 2}} }
  \left(\begin{array}{c}
    L_n^{(g-\frac{1}{2})}(x^2)
  \\ \\
  \mp \displaystyle{\frac{x}{\sqrt{n}}} \,
   L_{n-1}^{(g+\frac{1}{2})}(x^2)
  \end{array}\right)
\end{equation}
respectively. }

\smallskip

\item{For {$-1/2 < g < 1/2$}, it can be seen from (\ref{phi1-U}),
(\ref{phi2-U}) and (\ref{LI-sol-1}) that $\Phi_\lambda(x)\in
\mathbf{L_2}(0,1)$, $\forall \lambda \in \mathbb{C}$. This means
that, for these values of $g$, every complex number is an
eigenvalue of $Q_{+}^\dagger$ with degeneracy one. For example,
the eigenfunction of $Q_{+}^\dagger$ corresponding to $\lambda=i$
is given by
\begin{equation}\label{Phi+}\begin{array}{c}
  \Phi_{\lambda=i}(x)=\Phi_{+}(x)=
  \left( \begin{array}{c}
    \phi_{+,1} \\
    \phi_{+,2}
  \end{array} \right) = \\ \\
  = x^g\, e^{{- {x^2/ 2}} }
  \left(\begin{array}{c}
    U\left(\frac{1}{2},g+\frac{1}{2},x^2\right)
  \\ \\
  -\frac{i}{\sqrt{2}}\, x\, U\left(\frac{3}{2},g+\frac{3}{2},x^2\right)
  \end{array}\right),
\end{array}
\end{equation}
while the eigenfunction corresponding to $\lambda=-i$ is given by
its complex conjugate,
\begin{equation}\label{Phi-}
  \Phi_{\lambda=-i}(x)=\Phi_{-}(x)=\Phi_{+}(x)^*
\end{equation}
(since the coefficients in the differential operators in Eq.\
(\ref{eigen-problem}) are real). }

\smallskip

\item{ Finally, for $g\leq -1/2$, it can be seen that
$\Phi_\lambda(x)\notin \mathbf{L_2}(0,1)$ unless $-\lambda^2/2 =
g-\frac{1}{2}-n$, with $n=0,1,2,\dots$ In this case, taking into
account the Kummer transformation (see \cite{A-S}, page 505),
\begin{equation}\label{Kummer-transform}
  U(1-n-b,2-b,z)=z^{b-1}\,U(-n,b,z),
\end{equation}
and Eq.\ (\ref{U-poly}), we have $\phi_1(x)\sim x^{1-g}$ and
$\phi_2(x)\sim x^{-g}$ for $0<x\ll 1$. Therefore, in this region
$Q_{+}^\dagger$ has a symmetric real spectrum given by the
(degeneracy one) eigenvalues
\begin{equation}\label{spectrum-2}
 \lambda_{\pm,n}= \pm \sqrt{2 n+1-2g},\quad n=0,1,2,\dots
\end{equation}
corresponding to the eigenfunctions
\begin{equation}\label{eigenfunctions-2}\begin{array}{c}
  \Phi_{\pm,n}= (-1)^n \, n! \times \\ \\
  x^{-g}\, e^{{- {x^2/ 2}} }
  \left(\begin{array}{c}
   x\, L_n^{(\frac{1}{2}-g)}(x^2)
  \\ \\
  \mp \sqrt{n+\frac{1}{2}-g} \,
   L_{n}^{(-g-\frac{1}{2})}(x^2)
  \end{array}\right).
\end{array}
\end{equation}
Notice that no eigenvalue vanishes for these values of the
coupling.}

\end{itemize}

\smallskip

These results will be employed in the next Section to determine
the self-adjoint extensions of $Q_{+}$.

\section{Self-adjoint extensions of $Q_{+}$} \label{Sec-SAEQ+}

According to von Neumann's theory \cite{R-S}, to construct the
self-adjoint extensions of $Q_{+}$ we must take into account the
different behaviors of $Q_{+}^\dagger$, described in the previous
Section.

\subsection{For $|g|\geq 1/2$ the operator $Q_{+}$ is
essentially self-adjoint}

As seen in Section \ref{spectrum}, the {\it deficiency indices}
\cite{R-S} of $Q_{+}$, defined as the dimensions of the deficiency
subspaces $\mathcal{K}_\pm$,
\begin{equation}\label{defi-indi}
  n_\pm := {\rm dim}\ {\rm Ker}(Q_{+}^\dagger  \mp i ),
\end{equation}
vanish for $|g| \geq 1/2$. This means that $Q_{+}$ is {\it
essentially self-adjoint} \cite{R-S} in these regions of the
coupling, admitting there a unique self-adjoint extension given by
$Q_{+}^\dagger$ (which, in this case, is itself a self-adjoint
operator).

\smallskip

The corresponding self-adjoint extension of the Hamiltonian in
Eq.\ (\ref{susy-ham}) is given by \cite{teor-vN}
\begin{equation}\label{SAE-H-ESA}
  \overline{H}=(Q_{+}^\dagger)^2,
\end{equation}
where the operator composition in the right hand side is possible
in the dense domain
\begin{equation}\label{domain-P2}
  \mathcal{D} ( \overline{H} ) =
  \left\{ \psi \in \mathcal{D}(Q_{+}^\dagger):
  Q_{+}^\dagger \psi \in \mathcal{D}(Q_{+}^\dagger) \right\}.
\end{equation}
Notice that every eigenfunctions of $Q_{+}^\dagger$, corresponding
to an eigenvalue $\lambda$, belongs to $\mathcal{D} \left(
\overline{H} \right)$. Therefore, it is also an eigenfunction of
$\overline{H}$ with eigenvalue $E=\lambda^2$. So, we have:

\begin{itemize}

\item {For $g\geq 1/2$, the eigenfunctions of $\overline{H}$ are
given in Eqs.\ (\ref{eigenfunctions-1-0}) and
(\ref{eigenfunctions-1}). Notice that there is a unique zero mode,
while the positive eigenvalues of $\overline{H}$,
\begin{equation}\label{eigen-H-1}
  E_n=2\,n, \quad n=1,2,3,\dots
\end{equation}
are doubly degenerate (see Eq.\ (\ref{spectrum-1})). One can add
and subtract the corresponding eigenfunctions in Eq.\
(\ref{eigenfunctions-1}) to get {\it bosonic} and {\it fermionic}
states (with only the upper and lower component non vanishing
respectively). For these values of the coupling, the Witten index
is $\Delta = 1$ and the SUSY is manifest \cite{Witten}.}

\item {For $g\leq -1/2$, the eigenfunctions of $\overline{H}$ are
given in Eq.\ (\ref{eigenfunctions-2}). Notice that there is no
zero mode. Once again, the positive eigenvalues of $\overline{H}$,
\begin{equation}\label{eigen-H-2}
  E_n=2\,n+1-2\,g\geq 2, \quad n=0,1,2,\dots
\end{equation}
are doubly degenerate (see Eq.\ (\ref{spectrum-2})), and the
eigenfunctions can be combined to get bosonic and fermionic
states. For these values of $g$, the SUSY is spontaneously broken
and the Witten index is $\Delta = 0$ \cite{Witten}.}

\end{itemize}

\smallskip

\subsection{For $|g|< 1/2$ the operator $Q_{+}$ is not
essentially self-adjoint} \label{non-ESA}

On the other hand, according to Eqs.\ (\ref{Phi+}) and
(\ref{Phi-}) in Section \ref{spectrum}, for $-1/2 < g < 1/2$ the
deficiency indices are $n_\pm = 1$. In this region $Q_{+}$ admits
a one parameter family of self-adjoint extensions, $Q_{+}^\gamma$,
which are in a one-to-one correspondence with the isometries from
$\mathcal{K}_+$ onto $\mathcal{K}_-$  \cite{R-S}, characterized by
\begin{equation}\label{isometries}
  \mathcal{U}(\gamma) \Phi_+(x):= e^{2 i \gamma} \Phi_-,
  \quad \gamma \in [0,\pi),
\end{equation}
with $\Phi_+$ and $\Phi_-$ given in Eqs.\ (\ref{Phi+}) and
(\ref{Phi-}) respectively.

\smallskip

The self-adjoint operator $Q_{+}^\gamma$ is the restriction of
$Q_{+}^\dagger$ to a dense subspace
\begin{equation}\label{dom-Pgamma}
  \mathcal{D}(Q_{+}^\gamma)\subset
  \mathcal{D}(Q_{+}^\dagger ) =
  \mathcal{D}(\overline{Q}_{+})\oplus\mathcal{K}_+
  \oplus\mathcal{K}_-
\end{equation}
(here, $\overline{Q}_{+}$ is the closure of $Q_{+}$ \cite{R-S}),
composed by those functions which can be written as
\begin{equation}\label{func-dom-Pgamma}
  \Phi =\left( \begin{array}{c}
  \phi_1 \\
  \phi_2
\end{array}\right) =
\overline{\Phi}_0 + c \left( \Phi_+ + e^{2 i \gamma} \Phi_-
  \right),
\end{equation}
with $\overline{\Phi}_0 = \left(  \begin{array}{c}
  \phi_{0,1} \\
  \phi_{0,2}
\end{array} \right) \in \mathcal{D}(\overline{Q}_{+})$,
and the constant $c \in \mathbb{C}$.

Obviously, we have
\begin{equation}\label{accion-Pgamma}
  Q_{+}^\gamma \Phi = Q_{+}^\dagger  \overline{\Phi}_0 + i c \left(
  \Phi_+ - e^{2 i \gamma} \Phi_- \right),
\end{equation}
with $Q_{+}^\dagger $ given in Eq.\ (\ref{P+-def}).

\bigskip

Equation (\ref{func-dom-Pgamma}) completely characterizes the
behavior near the origin of the functions $\Phi \in
\mathcal{D}\left(Q_{+}^\gamma\right)$. As we will see, it also
allows to determine the spectrum of $Q_{+}^\gamma$.

Indeed, in Appendix \ref{closure} we have worked out the domain of
the closure of $Q_{+}$, $\mathcal{D}(\overline{Q}_{+})$, showing
that
\begin{equation}\label{phi0}
  \phi_{0,1}(x)= { o}(x^g), \quad \phi_{0,2}(x)={ o}(x^{-g}),
\end{equation}
for $x\rightarrow 0^+$. On the other side, from Eqs.\
(\ref{phi-eigen}), (\ref{phi1-U}), (\ref{phi2-U}) and
(\ref{LI-sol-1}), one can easily see that the components of any
eigenfunction $\Phi_\lambda$ of $Q_{+}^\dagger $ behave as
\begin{equation}\label{eigen-asymp}
  \begin{array}{c}
    \phi_1(x) = \displaystyle{\frac{\Gamma\left(
    \frac12 - g\right)}{\Gamma\left(
    \frac{1-\lambda^2}{2}-g\right)}}\ x^g + O(x^{1-g}),\\ \\
    \phi_2(x) = \displaystyle{\frac{\sqrt{2}}{\lambda} \
    \frac{\Gamma\left(
    \frac12 + g\right)}{\Gamma\left(
    -\frac{\lambda^2}{2}\right)}}\  x^{-g} + O(x^{1+g}).
  \end{array}
\end{equation}
Therefore, no eigenfunction of $Q_{+}^\dagger $ belongs to
$\mathcal{D}(\overline{Q}_{+})$.

\smallskip

Consequently, it is the contributions of $\Phi_\pm$ in Eq.\
(\ref{func-dom-Pgamma}) which determine the spectrum of
$Q_{+}^\gamma$. In fact, consider the limit
\begin{equation}\label{limit-ratio-phi}
  \lim_{x\rightarrow 0+} \frac{x^{-g}\, \phi_1(x)}{x^{g}\,
  \phi_2(x)}=
    \displaystyle{ \frac{\lambda}{\sqrt{2}} \,
    \frac{\Gamma\left(-\frac{\lambda^2}{2}\right)}
    {\Gamma\left(
    \frac{1-\lambda^2}{2}-g\right)}
    \frac{\Gamma\left(\frac12 - g\right)}
    {\Gamma\left(\frac12 + g\right)}   }.
\end{equation}
For a non vanishing $c$ in the right hand side of Eq.\
(\ref{func-dom-Pgamma}), this limit must coincide with
\begin{equation}\label{limit-ratio-phi+}
  \lim_{x\rightarrow 0+} \frac{\Re\left\{e^{-i \gamma}\, x^{-g}\,
   \phi_{+,1}(x)\right\}}
  {\Re\left\{e^{-i \gamma}\, x^{g}\, \phi_{+,2}(x)\right\}}=
    \displaystyle{- \sqrt{\frac{\pi}{2}} \,
    \frac{\cot(\gamma)}
    {\Gamma\left( 1-g\right)}
    \frac{\Gamma\left(\frac12 - g\right)}
    {\Gamma\left(\frac12 + g\right)}   },
\end{equation}
where Eq.\ (\ref{Phi-}) and  Eq.\ (\ref{eigen-asymp}) with
$\lambda \rightarrow i$ have been taken into account. Then, the
eigenvalues of $Q_{+}^\gamma$ (which are real) are the solutions
of the transcendental equation
\begin{equation}\label{trascendental}
    \displaystyle{ f(\lambda) :=
    \frac{\lambda \, \Gamma\left(-\frac{\lambda^2}{2}\right)}
    {\Gamma\left(
    \frac{1-\lambda^2}{2}-g\right)}
    = -\frac{\sqrt{\pi} \,\cot(\gamma)}
    {\Gamma\left( 1-g\right)} =: \beta(\gamma)}.
\end{equation}
Notice that $f(\lambda)$ is an odd function of $\lambda$, and
$-\infty \leq \beta(\gamma)< \infty$ for $0 \leq \gamma <\pi$.

\begin{figure}
    \epsffile{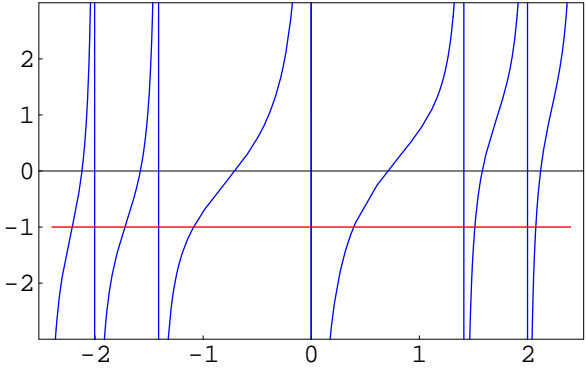}
    \caption{$f(\lambda):= \displaystyle{
    \frac{\lambda \, \Gamma\left(-\frac{\lambda^2}{2}\right)}
    {\Gamma\left(\frac{1-\lambda^2}{2}-g\right)}}$ for $g=1/4$,
    and $\beta(\gamma)\equiv -1$.} \label{fig1}
\end{figure}

The function $f(\lambda)$ in the left hand side of Eq.\
(\ref{trascendental}) has been plotted in Fig.\ 1,  for a value of
the coupling $g=1/4$. The eigenvalues of $Q_{+}^\gamma$ are
determined by the intersections of the graphic of $f(\lambda)$
with the horizontal line corresponding to the constant
$\beta(\gamma)$ (taken equal to $-1$ in the figure). As stressed
in Section \ref{spectrum}, the eigenvalues are non degenerate. The
eigenfunctions are obtained by replacing these eigenvalues in
Eqs.\ (\ref{phi-eigen}), (\ref{phi1-U}) and (\ref{phi2-U}).

\smallskip

It can be easily seen that, in general, the spectrum is non
symmetric with respect to the origin. The exceptions are the
self-adjoint extensions corresponding to $\gamma=0$
($\beta=-\infty$) and $\gamma=\pi/2$ ($\beta=0$). Indeed, the
condition $f(-\lambda)=f(\lambda)$ for a non vanishing $\lambda$
requires that
\begin{equation}\label{cond-sim}
  \frac{1}{\Gamma\left({-\frac{\lambda^2}{2}}\right)
  \Gamma\left( \frac{1-\lambda^2}{2}-g \right)} =0,
\end{equation}
whose solutions (see Fig.\ 1) correspond to the intersections with
the constant $\beta=-\infty$,
\begin{equation}\label{beta-infty}
  -\frac{\lambda^2}{2}=-n \Rightarrow
  \lambda_{\pm,n}=\pm\sqrt{2n},\quad n=1,2,3,\dots
\end{equation}
or the constant $\beta=0$,
\begin{equation}\label{beta-0}\begin{array}{c}
  \displaystyle{\frac{1-\lambda^2}{2}-g=-n \Rightarrow }\\ \\
  \lambda_{\pm,n}=\pm\sqrt{2n+1-2g},\quad n=0,1,2,\dots
\end{array}
\end{equation}

In particular, $Q_{+}^{\gamma=0}$ is the only self-adjoint
extension having a zero mode. For $0<\gamma<\pi$, the eigenvalues
are contained between contiguous asymptotes of $\Gamma\left(
-\frac{\lambda^2}{2} \right)$,
\begin{equation}\label{eigen-cotas}
  \sqrt{2 n}<\left| \lambda_{\pm,n} \right|<\sqrt{2(n+1)}.
\end{equation}

\bigskip

Now, for a given $Q_{+}^\gamma$, with $\gamma \in [0,\pi)$, we get
the self-adjoint extension of the Hamiltonian defined by
\cite{teor-vN}
\begin{equation}\label{SAE-H}
  H_\gamma =   (Q_{+}^\gamma)^2 \equiv\left.
  (Q_{+}^\dagger )^2 \right|_{\mathcal{D} \left( {H}_\gamma
  \right)},
\end{equation}
where the operator composition on the right hand side is the
restriction of $(Q_{+}^\dagger )^2$ to the dense subspace
\begin{equation}\label{domain-P-gamma-2}
  \mathcal{D} \left( {H}_\gamma \right) =
  \left\{ \psi \in \mathcal{D}\left(Q_{+}^\gamma\right):
  Q_{+}^\dagger  \psi \in \mathcal{D}\left(Q_{+}^\gamma\right) \right\}.
\end{equation}
This domain includes, in particular, all the eigenfunctions of
$Q_{+}^\gamma$, which are then also eigenvectors of $H_\gamma$:
\begin{equation}\label{eigen-H-gamma}
  Q_{+}^\gamma \Phi_\lambda = \lambda \Phi_\lambda \Rightarrow
  H_\gamma \Phi_\lambda = \lambda^2 \Phi_\lambda.
\end{equation}
Notice that, except for the special values $\gamma=0,\pi/2$, the
spectrum of ${H}_\gamma$ is non degenerate.

\smallskip

Three cases can be distinguished:

\begin{itemize}

\item{For $\gamma=0$ ($\beta=-\infty$) we get the only
self-adjoint extension of $H$ having a (non degenerate) zero mode.
The corresponding eigenfunction is also given by Eq.\
(\ref{eigenfunctions-1-0}). From Eq.\ (\ref{beta-infty}), it
follows that the non vanishing eigenvalues of $H_{\gamma=0}$ are
doubly degenerate,
\begin{equation}\label{eigen-gamma-0}
   E_{\pm,n} = \left( \lambda_{\pm,n} \right)^2 = 2n, \quad
  n=1,2,3,\dots
\end{equation}
We can take linear combinations of the corresponding
eigenfunctions, $\Phi_{\pm,n}$ (given by Eq.\
(\ref{eigenfunctions-1}), with $|g|<1/2$), to get linearly
independent states with only one non vanishing component.

Therefore, the conditions imposed on the functions in $\mathcal{D}
\left(Q_{+}^{\gamma=0}\right)$ by Eq.\ (\ref{func-dom-Pgamma})
with $\gamma=0$ give rise to a manifestly supersymmetric
self-adjoint extension of the Hamiltonian $H$. The Witten index is
in this case $\Delta=1$. }

\item{For $\gamma=\pi/2$ ($\beta=0$) we get a self-adjoint
extension of $H$ with no zero modes, and a doubly degenerate
spectrum. Indeed, from Eq.\ (\ref{beta-0}) it follows that the
self-energies of $H_{\gamma=\pi/2}$ are
\begin{equation}\label{eigen-gamma-pi}
  E_{\pm,n} = \left( \lambda_{\pm,n} \right)^2 =
  2n+1-2g, \quad n=0,1,2,\dots
\end{equation}
These eigenvalues are positive, since $1-2g>0$. The eigenfunctions
$\Phi_{\pm,n}$, whose expressions are given by Eq.\
(\ref{eigenfunctions-2}) with $|g|<1/2$, can be combined to get
bosonic and fermionic states.

In the present case, the conditions imposed on the functions in
$\mathcal{D} (Q_{+}^{\gamma=\pi/2})$ by Eq.\
(\ref{func-dom-Pgamma}) with $\gamma=\pi/2$ break the SUSY,
preserving the degeneracy of the spectrum. This gives a Witten
index $\Delta=0$. }

\item{For $\gamma\neq 0,\pi/2$ we get self-adjoint extensions of
$H$ with no zero modes and non degenerate spectra. The eigenvalues
of $H_\gamma$ (the square of those $\lambda$ solutions of Eq.\
(\ref{trascendental})) are all positive, and the corresponding
eigenfunctions are neither bosonic nor fermionic states (See Eqs.\
(\ref{phi1-U}) and (\ref{phi2-U})).

In this case, the condition imposed in Eq.\
(\ref{func-dom-Pgamma}) to select the domain of $Q_{+}^\gamma$
breaks not only the SUSY, but also the degeneracy of the spectrum.
The Witten index is $\Delta=0$. }

\end{itemize}

\bigskip

The  analysis performed in this Section should be compared with
the results obtained in \cite{DP}, where even and odd solutions
for a regularized version of this superpotential are worked out,
obtaining in the limit eigenfunctions belonging to the domains of
two different self-adjoint Hamiltonians, those corresponding to
$\gamma=0$ and $\gamma=\pi/2$.

\bigskip

{\small

\subsection{The $g=0$ case }
It is instructive to consider the $g=0$ case, in which the
superpotential (Eq.\ (\ref{W})) is regular at the origin, and the
functions in $\mathcal{D}(Q_{+}^\gamma)$ approach to constants for
$x\rightarrow 0^+$.

Indeed, if $g=0$, we have for the functions in the right hand side
of Eq.\ (\ref{func-dom-Pgamma}) (see Eqs.\ (\ref{lim0}) and
(\ref{eigen-asymp}))
\begin{equation}\label{BC-g=0}
  \begin{array}{c}
    \overline{\Phi}_0 (x)= o(x^0), \\ \\
    \Phi_+(x) + e^{2 i \gamma} \Phi_-(x) = 2^{3/2}\, e^{i \gamma}
    \left( \begin{array}{c}
      \sqrt{\frac{\pi}{2}}\, \cos\gamma \\
      -\sin \gamma
    \end{array} \right)+ O(x).
  \end{array}
\end{equation}

Therefore, the domain of $Q_{+}^\gamma$ can be characterized
simply by a local boundary condition of the form
\begin{equation}\label{BBCC-g=0}
  \Phi \in \mathcal{D}\left( Q_{+}^\gamma \right) \Rightarrow
  \left( \begin{array}{cc}
    \sin \gamma & \sqrt{\frac{\pi}{2}}\,\cos\gamma
  \end{array} \right) \cdot \left( \begin{array}{c}
    \phi_1(0) \\
    \phi_2(0)
  \end{array} \right)=0.
\end{equation}
The particular values $\gamma=0$ and $\gamma=\pi/2$ imply to
demand $\phi_2(0)=0$ and $\phi_1(0)=0$, respectively.

As discussed in Section \ref{non-ESA}, for $\gamma=0$ the SUSY is
manifest: There is a zero mode of $H_{\gamma=0}$,
\begin{equation}\label{zero-mode-g0}
  \Phi_0 = \left( \begin{array}{c}
    e^{-x^2/2} \\
    0
  \end{array} \right),
\end{equation}
and the eigenfunctions corresponding to the (doubly degenerate)
non vanishing eigenvalues, $E_{\pm,n}=2n\, , \  n=1,2,\dots$,
% \begin{equation}\label{En-gamma0}
%     E_{\pm,n}=2n\, , \quad n=1,2,\dots\,,
% \end{equation}
reduce to (see Eqs.\ (\ref{eigen-gamma-0}) and
(\ref{eigenfunctions-1}))
\begin{equation}\label{eigen-g0-gamma0}
  \Phi_{\pm,n}(x)=\frac{e^{-x^2/2}}{2^{2n}} \left( \begin{array}{c}
    H_{2n}(x) \\
    \pm 2 \sqrt{n} \, H_{2n-1}(x)
  \end{array} \right),
\end{equation}
where $H_n(x)$ are the Hermite polynomial. Notice that the lower
component and the first derivative of the upper component of the
eigenvectors vanish at the origin.

%\smallskip

For $\gamma=\pi/2$, the SUSY is spontaneouly broken: There are no
zero modes, and the eigenfunctions of $H_{\gamma=\pi/2}$
corresponding to the (doubly degenerate) non vanishing
eigenvalues, $E_{\pm,n}=2n+1$, $n=0,1,\dots$ reduce to (see Eq.\
(\ref{eigen-gamma-pi}) and (\ref{eigenfunctions-2}))
\begin{equation}\label{eigen-g0-gammapi}
  \Phi_{\pm,n}(x)=\frac{e^{-x^2/2}}{2^{2n+1}} \left( \begin{array}{c}
    H_{2n+1}(x) \\
    \mp \sqrt{4n+2} \, H_{2n}(x)
  \end{array} \right).
\end{equation}
In this case, the upper component and the first derivative of the
lower component of the eigenvectors vanish at the origin.

%\smallskip

For other values of the parameter $\gamma$, the SUSY is also
broken: There are no zero modes and the spectrum is non
degenerate, as previously discussed.

\smallskip

Therefore, we see that all except one of the possible local
boundary conditions at the origin defining a self-adjoint
supercharge $Q_{+}^{\gamma}$ (and a self-adjoint Hamiltonian
$H_\gamma$), Eq.\ (\ref{BBCC-g=0}), break the SUSY.

This should be compared with the results obtained in \cite{DP} for
the super half-oscillator, where the regularization employed for
the superpotential automatically leads to the eigenvalues and
eigenfunctions corresponding to the $\gamma=0$ case, Eqs.\
(\ref{zero-mode-g0}) and (\ref{eigen-g0-gamma0}), for which SUSY
is manifest.

}

\section{Discussion} \label{Discussion}

In the previous Sections we have seen how to choose suitable
domains to define self-adjoint extensions of the supercharge
$Q_{+}$, initially defined in the restricted domain
$\mathcal{C}^{\infty}_{0}(\mathbf{R^+\backslash \{0\}})$ as in
Eqs.\ (\ref{P-def}), (\ref{superA}) and (\ref{superAt}).

\smallskip

As stressed in Section \ref{setting}, $Q_{+}$ and $Q_{-}$ are
related by a unitary transformation (see Eq.\ (\ref{PP})). Then,
each self-adjoint extension of the first, $Q_{+}^{\gamma}$,
determines a self-adjoint extension of the second,
$Q_{-}^{\gamma}$, whose domain is obtained from
$\mathcal{D}(Q^{\gamma}_+)$ through the unitary transformation
$e^{{i\pi\sigma_3/4}}$,
\begin{equation}\label{dominio-SAE-Qmenos}
    \mathcal{D}(Q_-^{\gamma}) = \left\{
    \Psi : e^{-{i\pi\sigma_3/4}}\Psi \in
    \mathcal{D}(Q_+^{\gamma})\right\}
    = e^{\displaystyle{i\pi\sigma_3/4}} \left(
    \mathcal{D}(Q_+^{\gamma}) \right)\, .
\end{equation}
Consequently, $Q_{-}^{\gamma}$ is an equivalent representation of
the self-adjoint supercharge $Q_{+}^{\gamma}$, sharing both
operators the same spectrum.

\smallskip

Similarly, its square $(Q_{-}^{\gamma})^2$, defined on the dense
subspace \cite{teor-vN}
\begin{equation}\label{dominio-Qmenos-cuadrado}
    \begin{array}{c}
      \mathcal{D}\left((Q_-^{\gamma})^2 \right)
    =\left\{
    \Psi\in \mathcal{D}(Q_-^{\gamma})
     : Q_-^{\gamma} \Psi \in
    \mathcal{D}(Q_-^{\gamma})\right\} = \\ \\
    = e^{\displaystyle{i\pi\sigma_3/4}} \left(
    \mathcal{D}(H^{\gamma}) \right)\, ,
    \end{array}
\end{equation}
is an equivalent representation of the self-adjoint extension
$H^{\gamma}=(Q_+^{\gamma})^2 $ of the Hamiltonian $H$, initially
defined on $\mathcal{C}^{\infty}_{0}(\mathbb{R}^+\backslash
\{0\})$ as in Eq.\ (\ref{susy-ham}).

\smallskip

These equivalent representations of the Hamiltonian coincide only
if the domain $\mathcal{D}(Q^{\gamma}_+)$ is left invariant by the
unitary transformation $e^{{i\pi\sigma_3/4}}$, and this occurs
only for the particular self-adjoint extensions corresponding to
$\gamma = 0$ and $\gamma =\pi/2$ (extensions for which states can
be chosen to be \emph{bosons} or \emph{fermions}), as can be
easily seen from Eq.\ (\ref{limit-ratio-phi+}).

\smallskip

Consequently, the operator compositions
\begin{equation}\label{composicionQmasQmenos}
    (Q_+^{\gamma})^2,
    \quad (Q_-^{\gamma})^2,
    \quad Q_+^{\gamma}  Q_-^{\gamma}\qquad {\rm and} \qquad
    Q_-^{\gamma} Q_+^{\gamma}
\end{equation}
make sense in the same (dense) domain $\mathcal{D}(H^{\gamma})$
only for $\gamma = 0, \pi/2$, values of the parameter
characterizing self-adjoint extensions for which the $N=2$ SUSY
algebra is realized,
\begin{equation}\label{algebra-SUSY-0}
    \{ Q_+^{\gamma} , Q_-^{\gamma} \} = 0\,,\qquad
      H^{\gamma} =  (Q_+^{\gamma})^2 =
      (Q_-^{\gamma})^2\,.
\end{equation}

\smallskip

For other values of the parameter $\gamma$,
$\mathcal{D}(Q_+^{\gamma})$ is not left invariant by
$e^{{i\pi\sigma_3/4}}$, and there is no dense domain in the
Hilbert space where the self-adjoint operator compositions in Eq.\
(\ref{composicionQmasQmenos}) could be defined.

Therefore, for $\gamma \neq 0, \pi/2$ only one self-adjoint
supercharge can be defined in the domain of the Hamiltonian, and
the SUSY algebra reduces to the $N=1$ case,
\begin{equation}\label{algebra-SUSY-gamma}
      H^{\gamma} =  (Q_+^{\gamma})^2
\end{equation}
(or, equivalently, $(Q_-^{\gamma})^2 $).

\smallskip

At this point, it is worthwhile to remark that the double
degeneracy of the non vanishing eigenvalues of $H^{\gamma}$ with
$\gamma = 0, \pi/2$ is a consequence of the existence of a second
supercharge. Indeed, if
\begin{equation}
    Q^{\gamma}_+\Phi_\lambda=\lambda\Phi_\lambda\,,
\end{equation}
with $\Phi_\lambda\in \mathcal{D}(H^{\gamma})$ and $\lambda\neq
0$, then Eqs.\ (\ref{algebra-SUSY-0}) imply that
\begin{equation}\label{QmasQmenospsi}
    Q_+^{\gamma}(Q_-^{\gamma} \Phi_\lambda)
    =- Q_-^{\gamma}(Q_+^{\gamma} \Phi_\lambda)
    =- \lambda (Q_-^{\gamma} \Phi_\lambda)\, .
\end{equation}
Then, $Q_-^{\gamma} \Phi_\lambda$ ($\in \mathcal{D}(Q^{\gamma}_-)
\equiv \mathcal{D}(Q^{\gamma}_+)$) is a linearly independent
eigenvector of $Q^{\gamma}_+$ corresponding to the eigenvalue
$-\lambda$, since $Q_-^{\gamma} \Phi_\lambda \perp \Phi_\lambda$
and
\begin{equation}\label{Qmenos-phi-norma}
    \parallel Q_-^{\gamma} \Phi_\lambda \parallel^2
    =(\Phi_\lambda,(Q_-^{\gamma})^2
       \Phi_\lambda) = (\Phi_\lambda,H^{\gamma}
       \Phi_\lambda) = \lambda^2 \,
       \parallel \Phi_\lambda \parallel^2 \, \neq 0\,.
\end{equation}

\bigskip

In conclusion, we see that for a general self-adjoint extension of
the supercharge $Q_+^{\gamma}$ (and the corresponding extension of
the Hamiltonian, $H^{\gamma}$), the conditions the functions
contained in $\mathcal{D}(H^{\gamma})$ satisfy near the origin
prevent the $N=2$ SUSY, loosing one supercharge. Then, only the
$N=1$ SUSY algebra is realized, giving rise to a non symmetric
(and non degenerate) spectrum for the remaining supercharge, and a
non degenerate spectrum for the Hamiltonian. The remaining $N=1$
SUSY is spontaneously broken since there are no zero modes.

The only exceptions are those self-adjoint extensions
corresponding to $\gamma = 0$ and $\gamma = \pi/2$, for which the
$N=2$ SUSY algebra can be realized. In these two cases the
supercharges have a common symmetric (non degenerate) spectrum and
the excited states of the Hamiltonian are doubly degenerate.

For $\gamma = 0$, the (non degenerate) ground state of $H^{0}$ has
a vanishing energy and the SUSY is manifest, while for $\gamma=
\pi/2$ the (doubly degenerate) ground state of $H^{\pi/2}$ has
positive energy and the SUSY is spontaneously broken.

\bigskip

It is also worthwhile to point out that $N=2$ SUSY can be realized
only when the supercharge domain $\mathcal{D}(Q^{\gamma}_+)$ is
scale invariant. Indeed, consider a function $\Phi(x) \in
\mathcal{D}(Q^{\gamma}_+)$; under the scaling isometry
\begin{equation}\label{isometry}
    T_a \Phi(x):= a^{1/2} \Phi(a x)\,,
\end{equation}
with $a>0$, the limit in the left hand side of Eq.\
(\ref{limit-ratio-phi}) becomes
\begin{equation}\label{change-limit}
\begin{array}{c}
  \displaystyle{ \lim_{x\rightarrow 0+}
  \frac{x^{-g}\, (T_a\Phi)_1(x)}{x^{g}\,
  (T_a\Phi)_2(x)}}=
  \displaystyle{ \lim_{x\rightarrow 0+}
  \frac{x^{-g}\, a^{1/2}\, \phi_1(a x)}{x^{g}\,
  a^{1/2}\, \phi_2(a x)}}= \\ \\
   =a^{2 g} \displaystyle{ \lim_{y\rightarrow 0+}
   \frac{y^{-g}\, \phi_1(y)}{y^{g}\,
  \phi_2(y)}}=
  \displaystyle{- \sqrt{\frac{\pi}{2}} \,
    \frac{a^{2g} \cot(\gamma)}
    {\Gamma\left( 1-g\right)}
    \frac{\Gamma\left(\frac12 - g\right)}
    {\Gamma\left(\frac12 + g\right)}   }\, ,
\end{array}
\end{equation}
where Eq.\ (\ref{limit-ratio-phi+}) has been used in the last
step. This shows that $T_a \Phi(x)$ belongs to the domain of the
self-adjoint extension $Q^{\gamma_a}_+$ characterized by the
parameter $\gamma_a$ satisfying
\begin{equation}\label{gamma-a}
    \cot(\gamma_a)=a^{2 g} \cot(\gamma)\, .
\end{equation}
Obviously, $\gamma_a = \gamma$, $\forall\, a>0$, only for $\gamma
= 0, \pi/2$. For other values of $\gamma$ the conditions the
functions in $\mathcal{D}(Q^{\gamma}_+)$ satisfy near the origin
are not scale invariant.

\bigskip

Finally let us  stress that, as remarked in the Introduction, when
the formal expression of the Hamiltonian as a differential
operator is not essentially self-adjoint, additional information
is needed to identify  the self-adjoint extension which correctly
describes the properties of the physical system.

For the particular case under consideration we have seen that,
even though we have started from the {\em formal} $N=2$ SUSY
algebra of Eqs.\ (\ref{Ham-QQ}), (\ref{Q-Qtilde}), (\ref{nilpot}),
(\ref{superA}) and (\ref{superAt}), we  find a whole family of
self-adjoint extensions offering the possibility of having not
only spontaneously broken SUSY, but also a non-degenerate
Hamiltonian spectrum.

\bigskip

\noindent {\bf Acknowledgements:} The authors would like to thank
E.M. Santangelo and A. Wipf for useful discussions. They also
acknowledge support from Universidad Nacional de La Plata (Grant
11/X381) and CO\-NI\-CET, Argentina.

\appendix

\section{Closure of $Q_{+}$} \label{closure}

In this Section we will justify to disregard the contributions of
the functions in $\mathcal{D}(\overline{Q}_{+})$ to the
$x\rightarrow 0^+$ limit of the right hand side of Eq.\
(\ref{func-dom-Pgamma}). In fact, we will show that, near the
origin, $\overline{\Phi}_0(x)\in \mathcal{D}(\overline{Q}_{+})$
behaves as in Eq.\ (\ref{phi0}), for every $|g|<1/2$.

Since the graph of $Q_{+}$ is contained in the graph of
$Q_{+}^\dagger $, which is a closed set \cite{R-S}, it is
sufficient to determine the closure of the former. In so doing, we
must consider those Cauchy sequences
\begin{equation}\label{cauchy}
  \left\{ \Psi_n = \left( \begin{array}{c}
    \psi_{1,n} \\
    \psi_{2,n}
  \end{array} \right) \right\}_{n \in \mathbf{N}} \subset
  \mathcal{D}\left(Q_{+}\right) := \mathcal{C}_0^\infty
  (\mathbb{R^+}^+\backslash\{0\})
\end{equation}
such that $\left\{ Q_{+}\Psi_n \right\}_{n \in \mathbf{N}}$ are
also Cauchy sequences.

\smallskip

In this case, in particular, $\{ \psi_{1,n} \}_{n \in
\mathbb{N}}$, $\{ \psi_{2,n} \}_{n \in \mathbb{N}}$, $\{ A
\psi_{1,n} \}_{n \in \mathbb{N}}$ and $\{ \widetilde{A} \psi_{2,n}
\}_{n \in \mathbb{N}}$ are Cauchy sequences in ${\mathbf
L_2}(0,1)$, with $A$ and $\widetilde{A}$ given in Eqs.\
(\ref{superA}) and (\ref{superAt}) respectively.

Moreover, since $x$ is bounded in $[0,1]$, and the sum of
fundamental sequences is also fundamental, it follows that
$\left\{ \psi_{1,n}'(x)- \frac{g}{x}\, \psi_{1,n}(x) \right\}_{n
\in \mathbb{N}}$, and $\left\{ \psi_{2,n}'(x)+ \frac{g}{x}\,
\psi_{2,n}(x) \right\}_{n \in \mathbb{N}}$ are Cauchy sequences in
${ \mathbf L_2}(0,1)$.

\smallskip

On the other hand, we have $x^{\pm g}\in {\mathbf L_2}(0,1)$ for
any  $-1/2 < g <1/2$. Therefore,
\begin{equation}\label{cauchy2}
  \left\{ x^{-g} \left( \psi_{1,n}'(x)-\frac{g}{x}
  \psi_{1,n}(x) \right)\right\}_{n \in \mathbf{N}} = \left\{\left(
  x^{-g} \psi_{1,n}(x) \right)'\right\}_{n \in \mathbb{N}}
\end{equation}
and
\begin{equation}\label{cauchy3}
  \left\{ x^{g}  \left( \psi_{2,n}'(x)+\frac{g}{x}
  \psi_{2,n}(x) \right)\right\}_{n \in \mathbf{N}} = \left\{ \left(
  x^{g} \psi_{2,n}(x) \right)'\right\}_{n \in \mathbb{N}}
\end{equation}
are Cauchy sequences in ${\mathbf{L_1}}(0,1)$.

\smallskip

Now, taking into account that these functions vanish identically
in a neighborhood of the origin, one can see that $\left\{
x^{-g}\, \psi_{1,n}(x) \right\}_{n \in \mathbb{N}}$ and $\left\{
x^{g}\, \psi_{2,n}(x) \right\}_{n \in \mathbb{N}}$ converge
uniformly in $[0,1]$. Indeed, $\forall\, x \in [0,1] $ we have
\begin{equation}\label{conv-unif}\begin{array}{c}
  \Big|\, x^{-g} \left[ \psi_{1,n}(x)-\psi_{1,m}(x) \right] \Big|
  = \\ \\
  = \left| \int_0^x \left(y^{-g} \left[ \psi_{1,n}(y)-
  \psi_{1,m}(y) \right]\right)' \, dy\right| \leq \\ \\
  \leq \left\| \left(y^{-g} \psi_{1,n}(y)\right)' -
    \left(y^{-g} \psi_{1,m}(y)\right)' \right\|_{{\mathbf{L_1}}(0,1)}
    \rightarrow_{n,m\rightarrow \infty} 0,
\end{array}
\end{equation}
and similarly for the second sequence.

Consequently, there are two continuous functions,
$x^{-g}\phi_{0,1}(x)$ and $x^g \phi_{0,2}(x)$, which are the {\it
uniform} limits in $[0,1]$
\begin{equation}\label{limits}
  \begin{array}{c}
    x^{-g} \, \phi_{0,1}(x) =\displaystyle{\lim_{n\rightarrow
    \infty}} \
    x^{-g} \, \psi_{1,n}(x),
     \\ \\
    x^{g} \, \phi_{0,2}(x) = \displaystyle{\lim_{n\rightarrow
    \infty}}\
    x^{g} \, \psi_{2,n}(x).
  \end{array}
\end{equation}
In particular, we get
\begin{equation}\label{lim0}
  \begin{array}{c}
    \displaystyle{\lim_{x \rightarrow 0}}\  x^{-g} \,
    \phi_{0,1}(x) =0, \\ \\
    \displaystyle{\lim_{x \rightarrow 0}}\  x^{g} \,
    \phi_{0,2}(x) =0.
  \end{array}
\end{equation}

Moreover, the limit of the sequence in ${\mathbf{L_2}}(0,1)$ is
given by
\begin{equation}\label{lim-L2}
  \lim_{n \rightarrow \infty} \Psi_n =
  \overline{\Phi}_0=\left( \begin{array}{c}
  \phi_{0,1} \\
  \phi_{0,2}
\end{array} \right) .
\end{equation}
Indeed, taking into account that, for any $\varepsilon>0$,
\begin{equation}\label{inif-conv-lim}
  \Big|\, x^{-g}\left[ \psi_{1,n}(x) - \phi_{0,1}(x) \right] \Big|<
  \varepsilon, \quad \forall\, x\in [0,1],
\end{equation}
if $n$ is sufficiently large, it follows that
\begin{equation}\label{lim-L2-1}\begin{array}{c}
  \left\| \psi_{1,n} - \phi_{0,1} \right\|_{{\mathbf{L_2}}(0,1)}^2 = \\ \\
  = \int_0^1 x^{2g} \Big|\, x^{-g} \left(  \psi_{1,n}(x) -
  \phi_{0,1}(x) \Big)  \right|^2 <  \\ \\ < \varepsilon^2 \,
  \left\|x^g\right\|_{{\mathbf{L_2}}(0,1)}^2,
\end{array}
\end{equation}
and similarly for the lower component.

\smallskip

Eqs.\ (\ref{lim-L2}) and (\ref{lim0}) prove our assertion in Eq.\
(\ref{phi0}).

\bigskip

We will finally verify that the so obtained function
$\overline{\Phi}_0$ belongs to $\mathcal{D}({Q_{+}^\dagger })$.
Let $\rho_1(x)$ be the limit in ${\mathbf{L_1}}(0,1)$ of the
fundamental sequence given in Eq.\ (\ref{cauchy2}),
\begin{equation}\label{rho1}
  \rho_1(x)=\lim_{n \rightarrow \infty} {\left(x^{-g} \psi_{1,n}(x)\right)'} .
\end{equation}
Then, given $\varepsilon>0$, and $\forall\, x \in [0,1]$, we have
\begin{equation}\label{rho2}\begin{array}{c}
  \Big| \, x^{-g} \psi_{1,n}(x) - \int_0^x \rho_1(y) \, dy \Big| = \\ \\
  =\left| \int_0^x \left[ \left( y^{-g} \psi_{1,n}(y) \right)'
  -\rho_1(y) \right] dy \right|\leq \\ \\
  \leq \left\| \left( y^{-g} \psi_{1,n}(y) \right)'
  -\rho_1(y) \right\|_{{\mathbf{L_1}}(0,1)} < \varepsilon,
\end{array}
\end{equation}
if $n$ is large enough.

Since the uniform limit is unique, it follows from Eqs.\
(\ref{limits}) and (\ref{rho2}) that
\begin{equation}\label{rho-phi}
  \phi_{0,1}(x) = x^g \int_0^x \rho_1(y)\, dy,
\end{equation}
with $\rho_1 \in {{\mathbf{L_1}}(0,1)}$. Therefore,
$\phi_{0,1}(x)$ is an absolutely continuous function for $x>0$. A
similar conclusion is obtained for the lower component of
$\overline{\Phi}_0$.

\section{Spectral functions associated with $Q_{+}^\gamma$}
\label{spectral-functions}

\subsection{The graded partition function}\label{graded}

We will now consider the {\it graded partition function}
\cite{Cecotti,Bo,Smilga} of $H^\gamma$, defined as
\begin{equation}\label{grad-part}\begin{array}{c}
  Z_\gamma^F(T):= {\rm Tr}\left\{ \left(-1\right)^F e^{-T H^\gamma}
  \right\} =  \\ \\=
  \displaystyle{
  \sum_{\lambda_n} e^{-T \lambda_n^2} \frac{\left(
  \Phi_n, (-1)^F \Phi_n \right)}{\|\Phi_n\|^2} }.
\end{array}
\end{equation}
Subtracting the contribution of a possible zero mode we can write
\begin{equation}\label{grad-part-0}
    \widehat{Z_\gamma^F}(T):=\displaystyle{
  \sum_{\lambda_n\neq 0} \frac{e^{-T \lambda_n^2}}{\lambda_n}\
  \frac{\left( Q_{+}^\dagger  \Phi_n, (-1)^F \Phi_n
  \right)}{\|\Phi_n\|^2}},
\end{equation}
where
\begin{equation}\label{-1F}
  (-1)^F \left( \begin{array}{c}
  \phi_1   \\  \phi_2
\end{array} \right) = \left( \begin{array}{c}
  \phi_1   \\  - \phi_2
\end{array} \right).
\end{equation}
 Taking into account Eq.\ (\ref{P+-def}), and the fact that
the eigenfunctions are real, it is straightforward to get
\begin{equation}\label{grad-part-1}\begin{array}{c}
  \widehat{Z_\gamma^F}(T) =
  \displaystyle{ -
  \sum_{\lambda_n \neq 0} \frac{e^{-T \lambda_n^2}}
  {\sqrt{2} \lambda_n \|\Phi_n\|^2} \left[
  \phi_{n,1}(x)\phi_{n,2}(x)\right]_{x=0^+} =}\\  \\
  =  \displaystyle{ \frac{1}{2}
  \sum_{\lambda_n \neq 0} \frac{\Gamma\left( \frac{1}{2}+g \right)
  \Gamma\left( \frac{1}{2}- g \right)e^{-T \lambda_n^2}}
  {\Gamma\left(1- \frac{\lambda_n^2}{2}\right)
  \Gamma\left( \frac{1-\lambda_n^2}{2}-g \right)
  \|\Phi_n\|^2}},
\end{array}
\end{equation}
where the behavior of the functions in $\mathcal{D}\left(
H^{\gamma} \right)$ near the origin (see Eq.\ (\ref{eigen-asymp}))
has been taken into account in the last step.

We see that $Z_\gamma^F(T)$ depends on $\gamma$ though the
spectrum of $Q_{+}^\gamma$ and, in general, also depends on $T$.
But it can be shown that $Z_\gamma^F(T)$ is independent of $T$,
and coincide with the Witten index, for the particular values
$\gamma=0,\pi/2$.

Indeed, for the eigenvalues of $Q_{+}^{\gamma=\pi/2}$, given in
Eq.\ (\ref{eigen-gamma-pi}), each term in the series
in the right hand side of Eq.\ (\ref{grad-part-1}) vanishes
because of the second $\Gamma$- function in the denominator. So,
since there are no zero mode, we get
\begin{equation}\label{ZF-pi}
  Z_{\gamma=\pi/2}^F(T)\equiv 0=\Delta_{\gamma=\pi/2}.
\end{equation}

On the other hand, for the eigenvalues of $Q_{+}^{\gamma=0}$ given
in Eq.\ (\ref{eigen-gamma-0}), every term in the series in Eq.\
(\ref{grad-part-1}) vanishes because of the first
$\Gamma$-function in the denominator. In this case, we get from
the zero mode in Eq.\ (\ref{eigenfunctions-1-0})
\begin{equation}\label{ZF0}
  Z_{\gamma=0}^F(T)= \frac{\left(
  \Phi_0, (-1)^F \Phi_0 \right)}{\|\Phi_0\|^2}
  = 1 = \Delta_{\gamma=0}.
\end{equation}

For other values of $\gamma$, $Z_\gamma^F(T)$ vanishes
exponentially with $T$ (since there are no zero modes),
reproducing the Witten index in the  $T\rightarrow \infty$ limit.

\subsection{The spectral asymmetry} \label{spectral-asym}

The spectrum behavior for a general self-adjoint extension
$Q_{+}^\gamma$, as shown in Fig.\ \ref{fig1}, can be characterized
by the {\it spectral asymmetry} \cite{APS}
\begin{equation}\label{sepec-asymm}
  \eta(s):=\sum_{\lambda_{\pm,n} \neq 0} {\rm sign}
  \left( \lambda_{\pm,n}\right)\,
  \left| \lambda_{\pm,n} \right|^{-s}.
\end{equation}
Since $\left|\lambda_{\pm,n}\right| \sim \sqrt{n}$ (see Eq.\
(\ref{eigen-cotas})), Eq.\ (\ref{sepec-asymm}) defines an analytic
function on the open half plane $\Re(s)>2$.

For the particular values $\beta=-\infty$ and $\beta=0$, it is
evident from Eqs.\ (\ref{beta-infty}) and (\ref{beta-0}) that
$\eta(s)$ identically vanishes for any $g \in (-1/2,1/2)$.

The spectral asymmetry can also be expressed as
\begin{equation}\label{eta-zeta}
  \eta(s)= \zeta_+(s,\beta) - e^{i \pi s} \zeta_-(s,\beta),
\end{equation}
where
\begin{equation}\label{zeta+-}\begin{array}{c}
   \zeta_+(s,\beta):= \displaystyle{
   \sum_{\lambda_{+,n}>0} \lambda_{+,n}^{-s}, }
   \\  \\
  \zeta_-(s,\beta):= \displaystyle{
  \sum_{\lambda_{-,n}<0} \lambda_{-,n}^{-s}.}
\end{array}
\end{equation}

From Eq.\ (\ref{trascendental}), it can be seen that (for finite
$\beta(\gamma)$) the eigenvalues of $Q_{+}^\gamma$ are the zeros
of the analytic entire function
\begin{equation}\label{holomorphic}
  F(\lambda,\beta) := \frac{\lambda}{\Gamma\left(
  \alpha-\frac{\lambda^2}{2}\right)} -
  \frac{\beta}{\Gamma\left( -\frac{\lambda^2}{2} \right)},
\end{equation}
where $\alpha=\frac{1}{2}-g$. Since these zeroes are real and
simple, we have the following integral representation:
\begin{equation}\label{zeta+int}\begin{array}{c}
    \zeta_+(s,\beta)=\displaystyle{\frac{1}{2 \pi i} \oint_{\mathcal{C}_+}
  \lambda^{-s} \frac{F'(\lambda,\beta)}{F(\lambda,\beta)} \, d\lambda
  =} \\ \\
  \displaystyle{ = -\frac{1}{2\pi} \, e^{i\pi s/2}
  \int_{-\infty+i 0}^{\infty+i 0}
  \mu^{-s} \frac{F'(e^{-i\pi/2} \mu,\beta)}{F(e^{-i\pi/2} \mu,\beta)}
  \, d\mu,  }
\end{array}
\end{equation}
where $\mathcal{C}_+$ encloses counterclockwise the positive
zeroes of $Q_{+}^{\gamma}$.

Moreover, since $F(e^{i \pi}|\lambda|,\beta)= e^{i \pi}
F(|\lambda|,e^{- i \pi }\beta)$, it follows that the negative
zeroes of $F(\lambda,\beta)$ are {\it minus} the positive zeros of
$F(\lambda,e^{- i \pi }\beta)$. Consequently,
\begin{equation}\label{zeta-}
  \zeta_-(s,\beta) = e^{- i \pi s} \zeta_+(s,e^{-i\pi}\beta).
\end{equation}

\smallskip

Taking into account that
\begin{equation}\label{FpsobreF}\begin{array}{c}
  \displaystyle{\frac{F'(-i \mu,\beta)}{F(-i \mu,\beta)}}
    =\displaystyle{\frac{\displaystyle{
    1+\mu^2 \left[\psi\left(\mu^2/2\right)-
    \psi\left(\alpha+\mu^2/2\right) \right]}}
    {\displaystyle{- i \mu \left[
    1-i\,\beta\, \frac{ \Gamma\left(\alpha+\mu^2/2\right)}
    {\mu\, \Gamma\left(\mu^2/2\right)}\right]}}\, -} \\ \\
  -i \mu \,\psi\left(\mu^2/2\right) = i\, \Big[ \Delta_1(\mu)
  + \Delta_2(\mu,\beta) \Big] +
  O\left( \mu^{-3} \right),
\end{array}
\end{equation}
with
\begin{equation}\label{Delta-mu}
    \begin{array}{c}
      \Delta_1(\mu) = \displaystyle{
  - \mu \, \log \left(\frac{\mu^2}{2}\right)
  +\frac{1}{\mu} }, \\ \\
  \Delta_2(\mu,\beta) = \displaystyle{
   \frac{2g}
  {\mu\left[1-i\frac{\beta}{\mu}\left(
  \frac{\mu^2}{2}\right)^{-g+1/2}\right]} },
    \end{array}
\end{equation}
we see the right hand side of Eq.\ (\ref{zeta+int}) converges to
an analytic function on the open half-plane $\Re(s)>2$, region
from which it can be meromorphycally extended to the left.

For example, taking into account that
\begin{equation}\label{rotando}
  \frac{F'(-i e^{i \pi}  \mu,\beta)}
  {F(-i e^{i \pi}  \mu,\beta)}= e^{i \pi}
  \frac{F'(-i \mu,e^{-i \pi}\beta)}
  {F(-i\mu,e^{-i \pi} \beta)},
\end{equation}
we can write
\begin{equation}\label{zeta+asympt}
    \begin{array}{c}
   -2\pi \, \zeta_+(s,\beta)= \\ \\
   = \displaystyle{
   -2\, \sin \left( \frac{\pi s}{2} \right)
  \int_{1}^{\infty}
  \mu^{-s}  \Delta_1(\mu)\, d\mu \, +
  }  \\ \\
  \displaystyle{
  i \int_{1}^{\infty}
  \mu^{-s} \left\{ e^{i \pi s/2} \Delta_2(\mu,\beta) -
  e^{- i \pi s/2} \Delta_2(\mu,e^{- i \pi}\beta)
   \right\} d\mu  }\\ \\
  \displaystyle{
  +\,  e^{i \pi s/2}\int_{1}^{\infty}
  \mu^{-s}
   \left\{
  \frac{F'(-i \mu,\beta)}{F(-i
  \mu,\beta)}
  -\right.}\\ \\
  \displaystyle{\left.
   -i\Big[ \Delta_1(\mu)
  + \Delta_2(\mu,\beta) \Big] \right\} d\mu -}
  \\ \\
  \displaystyle{
  - e^{-i \pi s/2} \int_{1}^{\infty}
  \mu^{-s}
   \left\{
   \frac{F'(-i \mu,e^{-i \pi}\beta)}{F(-i
  \mu,e^{-i \pi} \beta)}
  -\right.}\\ \\
  \displaystyle{\left.
   -i\Big[ \Delta_1(\mu)
  + \Delta_2(\mu,e^{-i \pi} \beta) \Big] \right\} d\mu +}
  \\ \\
   \displaystyle{
  + e^{i \pi s/2} \int_{e^{i \pi}}^{1}
  \mu^{-s} \, \frac{F'(-i \mu,\beta)}{F(-i \mu,\beta)}
   \, d\mu  },
\end{array}
\end{equation}
where the first integral in the right hand side converges for
$\Re(s)>2$, the second one converges for $\Re(s)>0$, the third and
fourth ones exist for $\Re(s)>-2$,  and the fifth one (evaluated
on a curve going from $-1$ to 1 on the upper open half-plane, near
the real axis) is an entire function of $s$.

For the analytic extension of the first term on the right hand
side of Eq.\ (\ref{zeta+asympt}) we have
\begin{equation}\label{an-ext-1}
    \begin{array}{c}
     I_1(s)= \displaystyle{
   -2\, \sin \left( \frac{\pi s}{2} \right)
  \int_{1}^{\infty}
  \mu^{-s}  \Delta_1(\mu)\, d\mu }
   = \\ \\
      \displaystyle{ =
   - 2 \sin \left(\pi s/2\right) \left[\frac{1}{s} -
   \frac{2}{{\left(  s-2  \right) }^2} +
   \frac{\log (2)}
   { s-2}
   \right] ,}
   \end{array}
\end{equation}
and for the second one (calling $x=\mu^{-2g}$)
\begin{equation}\label{an-ext-2}
      \begin{array}{c}
     I_2(s)= \displaystyle{\Re \left\{
   2 i\, e^{i \pi s/2} \int_{1}^{\infty}
   \mu^{-s} \, \Delta_2(\mu,\beta) \, d\mu \right\}}
   = \\ \\
      \displaystyle{ =-\Re \left\{
      2 i\, e^{i \pi s/2} \lim_{\mu\rightarrow\infty}
      \int_1^{\mu^{-2g}} \frac{x^{\frac{s}{2\,g}-1 }  \, dx }
    {1 - i \,2^{g- \frac{1}{2}   }\,
       \beta\,x } \right\} ,}
   \end{array}
\end{equation}
for $g\neq 0$, while $I_2(s)\equiv 0$ for $g=0$.

According to the sign of $g$, we straightforwardly get:
\begin{itemize}

  \item {For $g>0$,
  \begin{equation}\label{I2-gpos}
    \begin{array}{c}
      I_2(s)=\displaystyle{-\frac{4 g}{s}\,
       \sin \left(\frac{\pi s}{2}\right) -
       \frac{2^{g+3/2} g\, \beta}{s+2 g} \,
        \cos \left(\frac{\pi s}{2}\right)
    + } \\ \\\displaystyle{
     + 2^{2 g} \beta^2
     \int_0^1 x^{\frac{s}{2 g}+1 } \,
     \frac{\sin \left(\frac{\pi s}{2}\right)+
      2^{g-1/2} \beta\, x\,
      \cos \left(\frac{\pi s}{2}\right)}{1 + 2^{2g-1}
     \beta^2  x^2} \, dx },
    \end{array}
\end{equation}
where the last integral converges for $s>- 4\, g$. Notice the
pole\footnote{This singularity implies that the $\zeta$-function
of $Q_{+}^\gamma$,
\begin{equation}\label{zeta}
  \zeta(s,\beta)\equiv\zeta_+(s,\beta)+\zeta_-(s,\beta)=
  \zeta_+(s,\beta)+e^{- i \pi s} \zeta_+(s,e^{-i \pi}\beta)
\end{equation}
presents a simple pole at $s=-2 g$,
\begin{equation}\label{pole-zeta}
  \zeta(s,\beta) = \frac{2^{ g+{3}/{2} }
    \left( e^{2 i \pi g }  -1\right)
       g\,\beta \,\cos (g\,\pi )}
    { s+ 2 g } + O(s+ 2 g)^0.
\end{equation}
The residue, which depends on the self-adjoint extension through
$\beta$, vanishes only for the $g=0$ case, and for $\beta =0$
(with any value of $g\in(-1/2,1/2)$). This is another example of a
singular potential leading to self-adjoint extensions with
associated $\zeta$-functions presenting poles at positions which
do not depend only on the order of the differential operator and
the dimension of the manifold, as is the general rule valid for
the case of smooth coefficients (see \cite{FPW,FMPS,FMP}).
 } at $s=-2 g$. }

  \item {For $g<0$ and $\beta\neq 0$,
  \begin{equation}\label{I2-gneg}
    \begin{array}{c}
      I_2(s)=\displaystyle{-
      \frac{2^{-g+5/2} g}{\beta (s-2 g) }
      \cos \left(\frac{\pi s}{2}\right)
     + } \\ \\\displaystyle{
    +  \int_1^\infty x^{ \frac{s}{2 g}-2 }\,
    \frac{2 \beta\, x \sin\left(\frac{\pi s}{2}\right) -
    2^{-g+3/2} \cos \left(\frac{\pi s}{2}\right)}{ \beta \left[
    1 + 2^{2g-1}  \beta^2  x^2\right]} \,  dx
     },
    \end{array}
\end{equation}
where the last integral converges for $s> 4\, g=-4|g|$. Notice the
pole at $s=2 g = - |2 g|$. }

\end{itemize}

\smallskip

Notice that $\zeta_+(s,\beta)$ is analytic in a neighborhood of
the origin. From Eqs.\ (\ref{zeta+asympt}), (\ref{an-ext-1}),
(\ref{I2-gpos}) and (\ref{I2-gneg}) it is easy to get the first
term of the Taylor expansion of $ \zeta_+(s,\beta)$  around $s=0$,
\begin{equation}\label{zeta+s0}
    \begin{array}{c}
   -2\pi \, \zeta_+(s \approx 0,\beta)= - \pi +
   \displaystyle{\left\{
   \begin{array}{cr}
     -2\pi g, & g>0 \\ \\
     0 , & g\leq 0
   \end{array}
    \right\}
   +}\\ \\
  \displaystyle{
  +   \int_{1}^{\infty}
  \left[\frac{F'(-i \mu,\beta)}{F(-i \mu,\beta)}
  -\frac{F'(-i \mu,e^{-i \pi}\beta)}{F(-i \mu,e^{-i \pi}\beta)}
  \right]\, d\mu +}\\ \\
   \displaystyle{+ i\Big[\log F(-i,\beta)-\log F(i,\beta)\Big]
   + O(s)},
\end{array}
\end{equation}
where the remaining integral can be evaluated taking into account
that
\begin{equation}\label{gsg}
  \displaystyle{\frac{\Gamma\left(\frac 1 2 -g +
  \frac{\mu^{2}}{2}\right)}{\mu \, \Gamma\left(
  \frac{\mu^{2}}{2}\right)} =
  2^{g-1/2} \mu^{-2g}\left\{ 1+
  O\left({\mu^{-2}}\right) \right\}}.
\end{equation}
We get
\begin{equation}\label{zeta+en0}
   \zeta_+(s = 0,\beta)=
   \left\{
   \begin{array}{l}
   \displaystyle{
     g, \quad 0<g < 1/2 , }\\ \\
     \displaystyle{ -
     \frac{1}{ \pi}\,
     \arctan\left( \frac{\beta}{\sqrt{2}}\right) ,\quad g=0,} \\ \\
     \displaystyle{
     -\frac 1 2 \, {\rm sign}(\beta) , \quad -1/2 < g<0.}
   \end{array}
    \right.
\end{equation}

\smallskip

Therefore, from Eqs.\ (\ref{eta-zeta}), (\ref{zeta-}) and
(\ref{zeta+en0}), it is straightforward  to get for the spectral
asymmetry of $Q_{+}^\gamma$ at $s=0$
\begin{equation}\label{eta+s0}
    \begin{array}{c}
   \eta(s = 0)= \left. \Big[
   \zeta_+(s ,\beta) -
   \zeta_+(s ,e^{-i\pi}\beta)\Big]\right|_{s=0}
   = \\ \\
   = \displaystyle{\left\{
   \begin{array}{l}
   \displaystyle{
     0 , \quad 0<g < 1/2 ,\ \forall\, \beta, }\\ \\
     \displaystyle{
     -\frac{2}{\pi} \arctan
     \left(\frac{\beta}
     {\sqrt{2}}\right), \quad g=0
     ,\ \forall\, \beta \in \mathbb{R^+},}
     \\ \\
     \displaystyle{- {\rm sign}(\beta),
     \quad - 1/2 < g<0,\ \forall\, \beta \neq 0,-\infty .}
   \end{array}
    \right.}
\end{array}
\end{equation}

% ----------------------------------------------------------------

% ----------------------------------------------------------------
\end{document}